\documentclass[conference]{IEEEtran}
\usepackage{fancyhdr}
\IEEEoverridecommandlockouts
\usepackage{cite}
\usepackage{amsmath,amssymb,amsfonts}
\usepackage{algorithmic}
\usepackage[pdftex]{graphicx}
\graphicspath{{./images/}}
\ifCLASSOPTIONcompsoc
 \usepackage[caption=false,font=normalsize,labelfont=sf,textfont=sf]{subfig}
\else
 \usepackage[caption=false,font=footnotesize]{subfig}
\fi
\usepackage{textcomp}
\usepackage{xcolor}
\usepackage{url}
\usepackage{adjustbox}
\usepackage{hhline}
\usepackage{algorithm}
\usepackage{multirow}
\usepackage{makecell}
\usepackage{etoolbox}
\usepackage{booktabs}
\usepackage{pgfplots}
\pgfplotsset{compat=1.18}
\usepgfplotslibrary{groupplots}
\usepackage{enumitem}
\usepackage{tagging}
\usepackage{sc26repro}
\usepackage{xurl}
\definecolor{barred}{RGB}{240,194,132}
\definecolor{curveblue}{RGB}{128,116,200}
\pgfplotsset{
    repairline/.style={color=curveblue, mark=*, mark size=0.9pt, line width=0.9pt}
}
\usepackage{microtype}
\pgfplotsset{
  repairbarstop/.style={
    scale only axis, width=3.9cm, height=2.9cm,
    axis y line*=left, axis x line*=bottom,
    xmin=-0.7, xmax=9.7,
    xtick={0,3,6,9}, xticklabels=\empty,
    ybar, bar width=0.22cm,
    tick label style={font=\scriptsize},
    every axis plot post/.append style={fill=barred, draw=barred!72!black},
  },
  repairbarsx/.style={
    scale only axis, width=3.9cm, height=2.9cm,
    axis y line*=left, axis x line*=bottom,
    xmin=-0.7, xmax=9.7,
    xtick={0,3,6,9},
    xlabel={Repair round}, xlabel style={font=\scriptsize, yshift=2pt},
    ybar, bar width=0.22cm,
    tick label style={font=\scriptsize},
    every axis plot post/.append style={fill=barred, draw=barred!72!black},
  },
  repaircurve/.style={
    scale only axis, width=3.9cm, height=2.9cm,
    axis y line*=right, axis x line*=none,
    xtick=\empty,
    xmin=-0.7, xmax=9.7, ymin=0, ymax=105,
    tick label style={font=\scriptsize},
  },
}

\newcommand{\oursC}[1]{\textcolor{green!60!black}{\textbf{#1}}}
\newcommand{\lassiC}[1]{\textcolor{red}{#1}}

\newbool{haveRAG} 
\boolfalse{haveRAG} 

\newcommand{\todo}[1]{\textcolor{blue}{TODO: #1}}

\def\BibTeX{{\rm B\kern-.05em{\sc i\kern-.025em b}\kern-.08em
    T\kern-.1667em\lower.7ex\hbox{E}\kern-.125emX}}

\makeatletter
\newcommand{\linebreakand}{%
  \end{@IEEEauthorhalign}
  \hfill\mbox{}\par
  \mbox{}\hfill\begin{@IEEEauthorhalign}
}
\makeatother

\begin{document}
\title{More with Less: Efficient Parallel Scientific Code Translation via an Agent-Skill Workflow}


%

\author{\IEEEauthorblockN{Jianuo Shen\IEEEauthorrefmark{1},
Pengyi Xu\IEEEauthorrefmark{2} and
Bo Zhang\IEEEauthorrefmark{2}}
\IEEEauthorblockA{\IEEEauthorrefmark{1} \texttt{csjianuo.shen@mail.scut.edu.cn} \\
\IEEEauthorrefmark{1} Department of Computer Science \& Engineering, South China University of Technology, Guangzhou, Guangdong 510006}
\IEEEauthorblockA{\IEEEauthorrefmark{2} \texttt{\{vxu, bo.zhang\}@mtu.edu} \\
\IEEEauthorrefmark{2} Department of Computer Science, Michigan Technological University, Houghton, Michigan 49931} 
}

\maketitle
\thispagestyle{fancy}
\lhead{}
\rhead{}
\chead{}
\lfoot{\footnotesize{
\copyright 2026 IEEE. Personal use of this material is permitted. Permission from IEEE must be obtained for all other uses... Accepted for publication in the Proceedings of the SC’26 Workshops (AgenticAI4HPC 2026).
}}
\rfoot{}
\cfoot{}
\renewcommand{\headrulewidth}{0pt}
\renewcommand{\footrulewidth}{0pt}

\begin{abstract}
Translating parallel scientific codes between HPC programming frameworks is a critical but labor-intensive task. Existing LLM-based approaches rely on fixed sequential pipelines with undifferentiated repair strategies and correctness criteria that do not match scientific computing practice. We present an agent-skill framework that separates translation, compilation diagnosis, and execution validation into three autonomous agents coordinated by a verdict-driven orchestrator, routing each failure to a targeted repair skill rather than retrying blindly. An automated output check drives the self-correction loop, complemented by an offline LLM-based verification that exposes false positives escaping compiler and runtime checks. We evaluate the framework on 30 HeCBench kernels translated bidirectionally among CUDA, OpenMP, and SYCL with three open-source LLMs. Our framework attains the higher verified correctness rate in all 18 model and direction combinations over a state-of-the-art baseline while consuming up to 52.4\% fewer tokens, remaining consistent across models and directions without architectural modification.
\end{abstract}


\section{Introduction}
The ability to automatically translate code between programming languages has long been a goal of software engineering research, and recent advances in large language models (LLMs) have brought it closer to practical reality. Modern LLMs such as GPT-4~\cite{openai2023gpt4}, DeepSeek Coder~\cite{guo2024deepseekcoder}, and Qwen2.5-Coder~\cite{hui2024qwen25coder} demonstrate strong generalization across programming languages, enabling translation pipelines that combine prompt engineering, compiler feedback, and iterative repair to achieve compelling results on general-purpose benchmarks~\cite{jiang2024llmcodesurvey,yao2023react}. These capabilities have sparked growing interest in applying LLMs to automate the translation of real-world codebases, reducing the manual effort required to port software across language ecosystems and opening new avenues for code modernization at scale.

Despite this progress, LLM-based code translation remains poorly understood in the domain of parallel scientific computing. General-purpose translation pipelines typically assess correctness through test case passing rates or code similarity metrics~\cite{chen2021codex,ren2020codebleu}, criteria that are adequate for software engineering tasks but insufficient for scientific computing workloads. Parallel scientific codes impose a stricter and more domain-specific notion of correctness: numerical outputs must be equivalent within floating-point tolerance across potentially billions of arithmetic operations, translated code must genuinely offload computation to the target device rather than silently falling back to CPU execution, and results must be deterministic across parallel runs~\cite{gopalakrishnan2017hpccorrectness,kaplan2026paracodex}. These requirements are difficult to satisfy precisely because the semantics of parallelism, memory hierarchy, and synchronization differ fundamentally across programming frameworks such as OpenMP~\cite{openmp2021}, CUDA~\cite{nvidia2024cudaguide}, and SYCL~\cite{khronos2023sycl}, a semantic gap that has no analog in conventional language-to-language translation and that general-purpose pipelines are not designed to bridge. Collectively, these failures remain invisible to evaluation methodologies that stop at compilation success or surface-level output matching, yet they are precisely the failures that matter most in scientific computing practice.

\ifbool{haveRAG}{
Several systems have recently addressed aspects of this problem, but share three recurring limitations. First, existing pipelines follow fixed sequential workflows with no dynamic routing: all errors are handled identically regardless of their root cause, leading to unnecessary LLM calls and poor token efficiency. LASSI~\cite{dearing2024lassi} exemplifies this pattern, achieving strong results on bidirectional OpenMP $\leftrightarrow$ CUDA translation with compiler and execution feedback, but applying the same repair strategy indiscriminately to every failure type. Second, automated semantic verification remains absent or superficial. CodeRosetta~\cite{tehranijamsaz2024coderosetta}, a purpose-built encoder-decoder transformer for C++ $\leftrightarrow$ CUDA and C++ $\leftrightarrow$ Fortran translation, reports 93\% functional correctness on a small sample of 30 kernels but does not assess numerical equivalence under floating-point arithmetic, the correctness criterion most critical for scientific workloads. LASSI similarly relies on manual output inspection rather than automated numerical validation. Third, domain knowledge is injected statically rather than retrieved adaptively. LASSI injects fixed-length API documentation regardless of the specific translation context, while systems such as UniPar~\cite{bitan2025unipar} and ParaCodex~\cite{kaplan2026paracodex} do not incorporate retrieval-augmented generation at all, limiting their ability to leverage the rich body of available HPC programming documentation and translation examples.
}{
Several systems have recently addressed aspects of this problem, but share three recurring limitations. First, existing pipelines follow fixed sequential workflows with no dynamic routing: all errors are handled identically regardless of their root cause, leading to unnecessary LLM calls and poor token efficiency. LASSI~\cite{dearing2024lassi} exemplifies this pattern, achieving strong results on bidirectional OpenMP $\leftrightarrow$ CUDA translation with compiler and execution feedback, but applying the same repair strategy indiscriminately to every failure type. Second, automated semantic verification remains absent or superficial. CodeRosetta~\cite{tehranijamsaz2024coderosetta}, a purpose-built encoder-decoder transformer for C++ $\leftrightarrow$ CUDA and C++ $\leftrightarrow$ Fortran translation, reports 93\% functional correctness on a sample of 30 kernels but does not assess numerical equivalence under floating-point arithmetic, the correctness criterion most critical for scientific workloads. LASSI similarly relies on manual output inspection rather than automated output validation. Third, existing pipelines are token-inefficient by design. LASSI and similar systems such as UniPar~\cite{bitan2025unipar} and ParaCodex~\cite{kaplan2026paracodex} apply the same LLM-based repair strategy regardless of error type, incurring unnecessary token consumption on failures that could be resolved with targeted, lightweight interventions. This indiscriminate retry behavior limits practical scalability and increases inference cost without a corresponding improvement in translation quality.
}

\ifbool{haveRAG}{
Inspired by the emerging paradigm of agent-skill decomposition in autonomous AI systems~\cite{yao2023react,xu2026agentskills,abuelsaad2024agente,li2026singleagentskills}, we propose an agentic workflow for parallel scientific code translation that addresses these limitations through principled separation of concerns. Our framework comprises three collaborating agents coordinated by an orchestrator: a \textit{TranslationAgent} that encapsulates code analysis, retrieval-augmented generation (RAG)~\cite{lewis2020rag}, memory-aware planning, and self-reviewing translation; a \textit{CompilationAgent} that combines compilation with structured error classification to dynamically route failures to the appropriate repair strategy; and an \textit{ExecutionAgent} that automatically validates numerical equivalence between source and translated outputs and diagnoses runtime failures including race conditions and incorrect GPU offloading. Rather than treating every error as a prompt for blind retry, our orchestrator reads the verdict from each agent and selects the repair skill best suited to the diagnosed failure mode, reducing unnecessary LLM calls and improving overall token efficiency. \todo{[PLACEHOLDER: evaluation scope sentence — to be finalized after evaluation results are confirmed.]}
}{
Inspired by the emerging paradigm of agent-skill decomposition in autonomous AI systems~\cite{yao2023react,xu2026agentskills,abuelsaad2024agente,li2026singleagentskills}, we propose an agentic workflow for parallel scientific code translation that addresses these limitations through principled separation of concerns. Our framework comprises three collaborating agents coordinated by an orchestrator: a \textit{TranslationAgent} that encapsulates code analysis, memory-aware planning, and self-reviewing translation; a \textit{CompilationAgent} that combines compilation with structured error classification to dynamically route failures to the appropriate repair strategy; and an \textit{ExecutionAgent} that automatically checks output correctness between source and translated outputs and diagnoses runtime failures including race conditions and incorrect GPU offloading. Rather than treating every error as a prompt for blind retry, our orchestrator reads the verdict from each agent and selects the repair skill best suited to the diagnosed failure mode, reducing unnecessary LLM calls and improving overall token efficiency. We evaluate the framework on 30 HeCBench~\cite{jin2023hecbench} kernels translated bidirectionally among OpenMP, CUDA, and SYCL with three recent open-source LLMs, measuring verified correctness and token consumption against a state-of-the-art baseline, and further audit the results under a stricter LLM-based complementary verification. Our results demonstrate that the agent-skill architecture achieves higher verified correctness than the baseline while reducing token consumption by up to 52.4\%, remaining consistent across three open-source LLMs and all six translation directions among CUDA, OpenMP, and SYCL without architectural modification.
}

This paper makes the following contributions:
\begin{itemize}
    \item We propose an agent-skill architecture for parallel scientific code translation that separates translation, compilation diagnosis, and execution validation into three autonomous agents. This separation enables verdict-driven routing that targets each repair skill at a specific failure mode, reducing unnecessary LLM calls while improving correctness through more precise self-correction.
    \item We introduce an automated output check as a first-class validation step, replacing manual output inspection with a quantitative correctness signal that drives a dedicated semantic repair loop, together with an LLM-based complementary verification that audits translated artifacts offline and exposes false positives escaping the output check.
    \item We demonstrate that our framework achieves higher translation correctness than LASSI on the same benchmark while reducing token consumption by up to 52.4\%, establishing that more capable behavior does not require more expensive prompting.
    \item We show that the framework generalizes across OpenMP, CUDA, and SYCL, supporting bidirectional translation between multiple HPC parallel programming frameworks without architectural changes.
\end{itemize}

The rest of this paper is organized as follows. Section~\ref{section:related_work} reviews related work. Section~\ref{section:design} describes the agent-skill architecture. Section~\ref{section:evaluation} presents experimental setup and results. Section~\ref{section:discussion} discusses findings and limitations. Section~\ref{section:conclusion} concludes and outlines directions for future work.

\section{Related Work} \label{section:related_work}
The rapid diversification of HPC hardware has made parallel scientific code portability one of the most pressing challenges in computational science. Leadership-class systems such as Frontier, Aurora, and El Capitan feature heterogeneous CPU-GPU architectures from different vendors, each requiring distinct parallel programming frameworks to achieve peak performance~\cite{shu2025hpcsurvey}. The current TOP500 list~\cite{top500} includes GPUs from NVIDIA, AMD, and Intel, driving a proliferation of parallel programming models including OpenMP, CUDA, SYCL, HIP~\cite{amd2016hip}, and Kokkos~\cite{trott2022kokkos3} that target overlapping but incompatible hardware ecosystems~\cite{davis2025gpumodels}. 
Manual porting is prohibitively expensive, with domain scientists estimating refactoring costs of upwards of eleven million dollars per 100,000 lines of code~\cite{matsuoka2022proxyapps}, and risks introducing subtle numerical and performance errors that are difficult to detect and reproduce. These pressures have driven growing interest in automated approaches to parallel scientific code translation, motivating the work surveyed below.

\subsection{LLM-Based Code Translation}
Automated code translation has a long history in software engineering. Early approaches relied on rule-based transpilers that applied hand-coded linguistic rules to map source language constructs to target language equivalents (as in SYCLomatic~\cite{syclomatic}, the production CUDA $\rightarrow$ SYCL migration tool), followed by statistical machine translation methods that applied phrase-based models originally designed for natural language to translate between programming languages such as Java and C\#~\cite{nguyen2013lexical,karaivanov2014phrase,nguyen2015divideconquer}. The emergence of neural models brought a significant shift: Roziere et al. introduced TransCoder~\cite{roziere2020transcoder}, an unsupervised transformer model that leverages masked language modeling, denoising autoencoding, and back-translation to translate between Python, Java, and C++ without requiring parallel training data.
Subsequent work improved upon this foundation by incorporating automated unit test execution to filter invalid translations~\cite{roziere2022transcoderst} and by injecting code structure through abstract syntax trees and compiler representations~\cite{szafraniec2023codetranslation}.
With the rise of large language models, prompt-based approaches have largely supplanted purpose-built translation models, enabling zero-shot and few-shot translation across a much wider range of language pairs without task-specific training~\cite{jiang2024llmcodesurvey}.
Compiler feedback and iterative repair loops have emerged as an effective complement to prompting, feeding compilation and execution errors back to the LLM for self-correction~\cite{chen2021codex}. Despite these advances, general-purpose translation systems assess correctness through code similarity metrics~\cite{ren2020codebleu} or test case passing rates~\cite{chen2021codex}, criteria that are insufficient for parallel scientific codes where numerical equivalence and genuine GPU offload are the relevant correctness standards.

Applying these advances to parallel scientific codes has motivated a dedicated line of work in the HPC community.
Lei et al.~\cite{lei2023hpc} introduced a dataset of manually curated OpenMP Fortran and C++ code pairs and demonstrated that fine-tuning LLMs on HPC-specific translation data significantly improves translation quality over general-purpose baselines.
LASSI~\cite{dearing2024lassi} introduced a self-correcting pipeline for bidirectional CUDA $\leftrightarrow$ OpenMP translation combining self-prompting, static domain knowledge injection, and iterative compiler and execution feedback, but follows a fixed sequential workflow that applies the same repair strategy to every failure type and relies on manual output inspection rather than automated output verification; its extension LASSI-EE~\cite{dearing2025lassiee} targets energy-aware refactoring, without numerical equivalence verification. CodeRosetta~\cite{tehranijamsaz2024coderosetta} proposed an unsupervised encoder-decoder transformer for C++ $\leftrightarrow$ CUDA and C++ $\leftrightarrow$ Fortran translation with training objectives tailored to parallel code semantics, but evaluates correctness on a small sample of kernels without assessing numerical equivalence under floating-point arithmetic. UniPar~\cite{bitan2025unipar} evaluated fine-tuning, hyperparameter optimization, and compiler-guided repair for translation among serial code, CUDA, and OpenMP, improving over off-the-shelf models while leaving numerical correctness unverified.
ParaCodex~\cite{kaplan2026paracodex} introduced profiling-guided agentic planning for serial and CUDA to OpenMP GPU offload translation, using hotspot analysis and explicit data movement planning to prevent performance regressions, but focuses on parallelization rather than bidirectional translation between established parallel programming frameworks.
Evaluation in this space has primarily relied on kernel-level benchmarks such as HeCBench~\cite{jin2023hecbench}, which provides over 500 computational kernels with CUDA, OpenMP, and SYCL implementations. Most recently, ParEval-Repo~\cite{davis2025parevalrepo} extended evaluation to repository-level HPC translation, finding that existing approaches largely fail beyond trivial kernel scale due to cross-file dependencies and build system complexity. Across these systems, three recurring limitations emerge: fixed sequential workflows with no dynamic error-type-aware routing, absence of automated correctness verification, and indiscriminate repair strategies that incur unnecessary token consumption regardless of failure type.

\subsection{Agentic Workflows for Software Engineering}
The application of LLM-based agents to software engineering tasks has grown rapidly, driven by the observation that complex software tasks require multi-step reasoning, tool use, and iterative self-correction that single-prompt approaches cannot provide.
ReAct~\cite{yao2023react} established the foundational paradigm of interleaving reasoning and acting in LLM-based agents.
SWE-agent~\cite{yang2024sweagent} demonstrated that agent-computer interfaces enabling LLMs to interact with terminals, editors, and test runners substantially outperform static prompt-based approaches on real-world GitHub issue resolution. AutoCodeRover~\cite{zhang2024autocoderrover} combined LLMs with AST-based program analysis to autonomously localize and repair bugs, while RepairAgent~\cite{bouzenia2025repairagent} showed that LLM-based agents with access to compilation and test execution feedback can navigate complex repair trajectories autonomously. These systems share a common architectural insight: separating high-level decision-making from low-level execution enables more robust workflows than monolithic prompt chains.

Skills as stateless execution units have gained attention for improving modularity and token efficiency. 
Rather than routing all tasks through a single LLM call or distributing them across communicating agents, skill-based architectures assign lightweight, targeted execution to specialized stateless components, reducing token usage compared to multi-agent alternatives~\cite{li2026singleagentskills}, while reserving expensive LLM reasoning for decisions that genuinely require it~\cite{xu2026agentskills}.
The advantage of dynamic, conditional execution over fixed sequential pipelines has also been demonstrated directly in code translation:
Gray et al.~\cite{gray2025workflowsvsagents} found that an agentic MATLAB $\rightarrow$ HDL translation pipeline with conditional tool invocation outperforms fixed sequential workflows by over 20\% points, with gains attributed to short prompts and on-demand tool use.
These agentic software engineering systems, however, assess correctness through test suite execution, criteria that do not transfer to parallel scientific codes where numerical equivalence, genuine GPU offload, and parallel determinism are the relevant correctness standards. Our work applies the agent-skill paradigm to this domain, introducing domain-specific correctness verification as a first-class component of the agentic workflow rather than a post-hoc evaluation step.

\section{System Design} 
\label{section:design}

\begin{figure*}[t]
    \centering
    \includegraphics[width=0.75\textwidth]{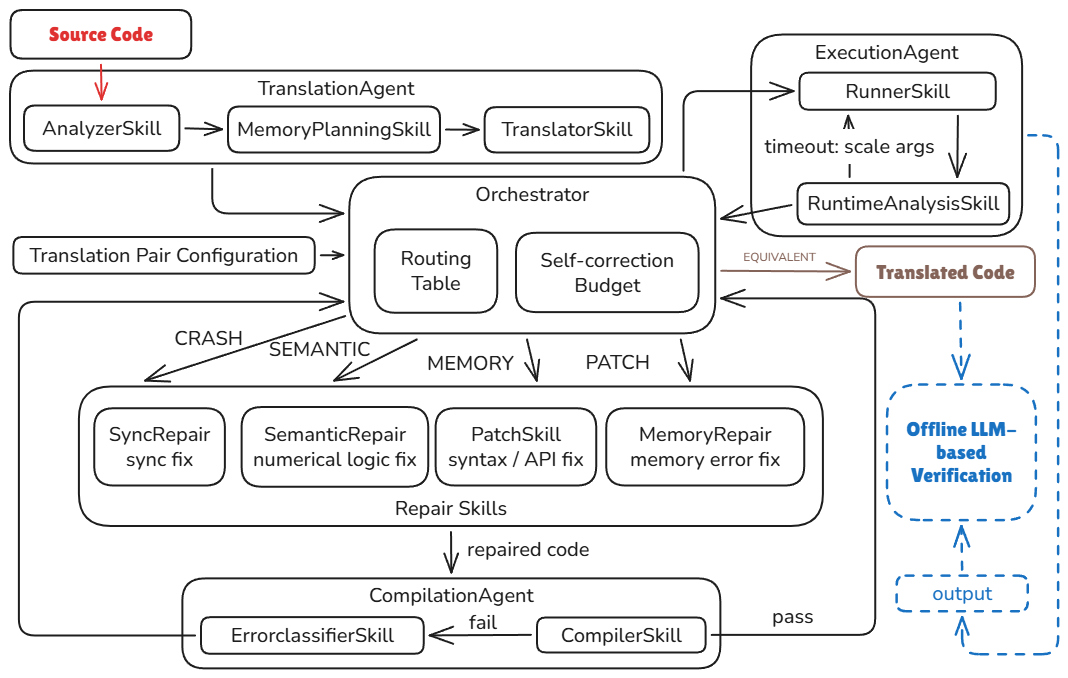}
    \vspace{-5pt}
    \caption{System overview. The orchestrator drives a verdict-based self-correction loop through the three agents, routing each diagnosed failure to a targeted repair skill until the translation passes compilation and output checking or the self-correction budget is exhausted.}
    \label{fig:system_design}
    \vspace{-5pt}
\end{figure*}

This section presents the design of our LLM-driven code translation framework. Translating parallel programs between programming models is rarely a one-shot task: LLM-generated code frequently fails to compile, and code that compiles may still produce incorrect results at runtime. Rather than treating translation as a single generation step, our framework organizes the task as a multi-agent system in which specialized agents collaborate through a central orchestrator and an iterative repair loop. An LLM is invoked for code generation, rewriting, analysis, and planning tasks. Meanwhile, in-loop verification is delegated to the real compiler and runtime behavior, complemented by an offline LLM-based complementary verification.

\subsection{Overview and Design Goals}
Our framework translates parallel programs between different parallel programming models through a multi-agent system coordinated by a central orchestrator. As shown in Figure \ref{fig:system_design}, the system consists of three agents: a \textit{TranslationAgent} generates the target-language code, a \textit{CompilationAgent} validates that the code compiles, and an \textit{ExecutionAgent} runs the compiled binary and checks correctness. Each agent is composed of a fixed set of skills that it executes in sequence: the \textit{TranslationAgent} runs code analysis, memory planning, and translation skills; the \textit{CompilationAgent} runs compiler invocation and error classification skills; and the \textit{ExecutionAgent} runs binary execution and runtime analysis skills. In addition, the framework provides specialized repair skills to handle distinct compilation and runtime failure modes.

The orchestrator drives the workflow through a verdict-based self-correction loop. After each agent or repair skill executes, the orchestrator records the returned verdict and consults a routing table that maps every verdict status to the next agent, a repair skill, or a terminal state. The loop terminates upon reaching a terminal verdict, exhausting a global step limit, or exceeding a unified self-correction budget that counts only repair invocations. This verdict-driven routing separates the concerns of code generation, failure classification and repair.

The overarching goal of the framework is to support translation between arbitrary parallel programming models. To make this practical, concrete translation logic is encapsulated in dynamically loaded skills rather than baked into the agent or orchestrator code, so adding a new source$\rightarrow$target translation pair requires only pair-specific skill implementations and prompts without modifying the system architecture.
The active pair is selected at runtime through a single configuration flag, which swaps the bound skill implementations, prompt templates, and compiler choices.

\begin{table*}[t]
    \centering
    \caption{Classification of compilation and runtime failures into verdict categories.}
    \vspace{-2pt}
    \label{tab:verdict-classification}
    \footnotesize
    \setlength{\tabcolsep}{4pt}
    \renewcommand{\arraystretch}{0.95}
    \begin{tabular}{@{}llp{5.4cm}ll@{}}
        \toprule
        \textbf{Agent} & \textbf{Skill} & \textbf{Trigger condition} & \textbf{Verdict} & \textbf{Routed to} \\
        \midrule
        \multirow{7}{*}{CompilationAgent}
            & \multirow{7}{*}{ErrorClassifierSkill}
            & Linker errors (undefined reference, multiple definition, \texttt{ld} failure)
                & \texttt{PATCH} & PatchSkill \\
            & & Leftover source-framework keywords (e.g., \texttt{\_\_host\_\_}, \texttt{\_\_device\_\_})
                & \texttt{PATCH} & PatchSkill \\
            & & Target-framework API or directive misuse (e.g., \texttt{sycl::}, \texttt{omp target}, \texttt{cudaMalloc})
                & \texttt{PATCH} & PatchSkill \\
            & & Genuine memory-allocation failures (out of memory, invalid device pointer)
                & \texttt{MEMORY} & MemoryRepair \\
            & & Missing headers or source files
                & \texttt{PATCH} & PatchSkill \\
            & & C++ syntax, type, or scope errors
                & \texttt{PATCH} & PatchSkill \\
            & & Empty compiler log (no actionable error)
                & \texttt{RETRANSLATE} & TranslationAgent \\
        \midrule
        \multirow{4}{*}{ExecutionAgent}
            & \multirow{4}{*}{RuntimeAnalysisSkill}
            & Exit code 0 and no \texttt{FAIL} marker in stdout
                & \texttt{EQUIVALENT} & Done (terminal) \\
            & & Exit code 0 but explicit \texttt{FAIL} marker in stdout
                & \texttt{SEMANTIC} & SemanticRepair \\
            & & Non-zero exit code
                & \texttt{CRASH} & SyncRepair \\
            & & Execution timeout; args auto-scaled and rerun
                & \texttt{REPAIR\_EXECUTION} & RunnerSkill \\
        \bottomrule
    \end{tabular}
    \vspace{-5pt}
\end{table*}

\subsection{Task State, Verdict, and Orchestrator}
\label{section:Task State, Verdict and Orchestrator}

The framework uses a shared \textit{TaskState} object to represent the evolving state of a single translation task as it moves through the orchestrator's self-correction loop. Initialized from the source program and pair configuration, it tracks the current translated code, the latest compilation and execution artifacts, and bookkeeping information such as step history and retry counters. A generic metadata bucket inside the \textit{TaskState} allows skills to share intermediate context, such as code summaries and memory plans. Both agents and skills mutate this state in place, which simplifies the communication protocol.

Agents and skills communicate through a structured Verdict object that encodes the outcome of the most recent step. The orchestrator invokes \textit{CompilationAgent} or \textit{ExecutionAgent}, which maps compiler log or runtime output to a verdict category according to a set of trigger conditions. The orchestrator then dispatches the verdict to the corresponding repair skill or back to \textit{TranslationAgent} via a fixed routing table. The mapping from trigger conditions to verdicts and their routing targets is given in Table~\ref{tab:verdict-classification}.

\begin{table*}[t]
  \centering
  \caption{Core content of skill prompts. Placeholders in braces are filled at runtime.}
  \vspace{-2pt}
  \label{tab:prompt-core}
  \scriptsize
  \setlength{\tabcolsep}{4pt}
  \renewcommand{\arraystretch}{0.95}
  \begin{tabular}{@{}p{0.15\textwidth} p{0.80\textwidth}@{}}
    \toprule
    \textbf{Skill} & \textbf{Core Prompt Content} \\
    \midrule
    AnalyzerSkill (shared) & System: You are a code analysis assistant. Describe what a program does in plain, framework-independent terms. User: Analyze the following \{source\_language\} code. Focus only on the real execution logic and numerical behavior. Include: what problem the program solves, inputs and expected outputs, computation stages, correctness criteria, and numerical concerns. Do NOT focus on \{target\_language\} API names, \{target\_language and runtime constructs\} or synchronization details. Keep the response under 250 words. \\
    MemoryPlanningSkill & System: You are an expert in \{target\_language\} parallelism. User: Given the following \{source\_language\} code, produce a memory plan for \{target\_language\} execution. List each buffer with: variable name, size, lifetime, and sharing behavior. Identify where \texttt{reduction}, \texttt{atomic}, \texttt{critical}, or \texttt{barrier} clauses are needed. Respond as a bullet list only. \\
    TranslatorSkill & [BEGIN CODE SUMMARY] \{code\_summary\} [END CODE SUMMARY] [BEGIN MEMORY PLAN] \{memory\_plan\} [END MEMORY PLAN] Generate new code to refactor the following parallelized \{source\_language\} program to instead use C++ code written with \{target\_language\} directives. \\
    PatchSkill & The following code was compiled with \{compiler\} \{flags\} and produced the following compile error: \{compile\_log\} Re-factor the above code with a fix to eliminate the stated error. \\
    SemanticRepair & System: You are a coding assistant that fixes numerical logic errors in \{target\_language\} code. User: The translated code was executed and produced incorrect or unexpected output. Translated stdout: \{run\_stdout\} Translated stderr: \{run\_stderr\} Identify and fix the numerical logic error. \\
    SyncRepair & System: You are an expert in \{target\_language\} parallel programming synchronization. User: The following \{target\_language\} code crashes or produces non-deterministic results (exit code \{run\_exit\_code\}, stderr: \{run\_stderr\}). Add missing barriers, atomic operations, or reduction clauses as needed. \\
    MemoryRepair & System: You are a coding assistant that fixes memory-related compilation errors in \{target\_language\} code. User: The following code was compiled with \{compiler\} \{flags\} and produced a compile error related to memory usage: \{compile\_log\} [BEGIN MEMORY PLAN] \{memory\_plan\} [END MEMORY PLAN] Fix the memory issue. \\
    \bottomrule
  \end{tabular}
  \vspace{-5pt}
\end{table*}

The Orchestrator implements the central control loop. It consumes verdicts, advances the pipeline to the next agent or repair skill, and enforces termination conditions.
The orchestrator also maintains the unified self-correction budget, configurable at construction time, whose cap on repair attempts applies across the entire system.
Forward progress through translation, compilation, and execution does not consume this budget; only repair invocations do. This separation ensures that successful translations are not penalized by the presence of repair machinery, while preventing runaway retry loops.

\subsection{Agent Skills and Repair Mechanisms}
\label{section:agent-skills}

Each agent in the system is composed of one or more core skills that encapsulate a single capability. Repair skills are managed directly by the orchestrator and invoked when a stage returns a failure verdict.

The \textit{TranslationAgent} runs three skills in sequence. First, the \textit{AnalyzerSkill} invokes an LLM to produce a framework-independent functional summary of the source program, describing its computation stages, inputs, outputs, and correctness criteria. The summary is stored in the shared state and later consumed by the translator. Then, the \textit{MemoryPlanningSkill} generates a memory and data-layout plan for the target execution model, mapping each buffer to its intended sharing behavior, lifetime, and synchronization requirements. Finally, the \textit{TranslatorSkill} synthesizes the complete target-language program by combining the source code with the functional summary and the memory plan. 

The \textit{CompilationAgent} contains two skills. The \textit{CompilerSkill} invokes the real target-language compiler and captures the compilation log and the resulting binary path. If compilation succeeds, the agent returns a pass verdict and control advances to the \textit{ExecutionAgent}; otherwise, the \textit{ErrorClassifierSkill} applies a set of trigger condition as shown in Table \ref{tab:verdict-classification} to the compiler log as presented in Section \ref{section:Task State, Verdict and Orchestrator}.

The \textit{ExecutionAgent} also contains two skills. The \textit{RunnerSkill} executes the compiled binary with test arguments and records \texttt{stdout}, \texttt{stderr}, and the exit code. The \textit{RuntimeAnalysisSkill} then maps the recorded output to a verdict category according to the trigger conditions in Table~\ref{tab:verdict-classification}. If execution times out, the \textit{RuntimeAnalysisSkill} automatically scales down numeric test arguments that exceed a configurable threshold. The scaling factor is halved across attempts, and the modified arguments are written back to the \textit{TaskState} before the binary is rerun with smaller inputs. This mechanism allows computationally intensive kernels to be validated without manual input tuning. And the total number of rescaling attempts is configurable and defaults to five.

Four repair skills handle the failure modes exposed by compilation and execution. Two address compilation failures: the \textit{PatchSkill} rewrites the current code to eliminate a reported compilation error, and the \textit{MemoryRepair} specializes in memory-related compilation errors, using the previously generated memory plan to correct allocation, mapping, or sharing mistakes. The other two repair skills address runtime failures: the \textit{SemanticRepair} fixes numerical or logical errors observed at runtime, such as incorrect loop bounds, missing reductions, or precision loss, and the \textit{SyncRepair} addresses crashes and non-deterministic behavior by inserting missing barriers, atomic operations, or synchronization constructs. Each repair skill invokes an LLM to generate corrected code and returns a repair-compilation verdict, sending the modified code back to the \textit{CompilationAgent} and closing the self-correction loop.

The prompts that drive these skills are summarized in Table~\ref{tab:prompt-core}. The \textit{AnalyzerSkill} uses a shared prompt across all translation pairs, while the \textit{TranslatorSkill} and repair skills use pair-specific prompts that follow a common structure but encode different target-model constraints.

\subsection{Compilation, Runtime, and LLM-based Verification}
\label{sec:verification}

The framework adopts an in-loop verification stage plus an LLM-based complementary verification. The in-loop verification stage relies on compiler and runtime feedback to validate translated programs and provide signals for the self-correction workflow. Compilation success is treated as the prerequisite for execution, while runtime behavior is evaluated through execution status and program outputs. A translation is accepted by this stage only when the generated binary exits with code zero and its output contains no explicit \texttt{FAIL} marker (a failure-indicating word in the program output), corresponding to the \texttt{EQUIVALENT} verdict in Table~\ref{tab:verdict-classification}.

Although compiler and runtime verification provide reliable and reproducible signals, they cannot capture all correctness issues in translated programs. A program may compile successfully and terminate normally while still failing to execute the intended workload, producing incomplete computations, or generating incorrect numerical results without explicit error messages. Such false positives can lead to an overestimation of translation success rates and obscure the actual capability of the translation framework.

To address these limitations, we introduce an LLM-based complementary verification. Unlike the compiler and runtime checks that participate directly in the self-correction workflow, the LLM verifier is applied offline to cached translation artifacts and aims to identify defects that escape compiler and runtime checks. The verifier receives the complete source program and execution outputs, and produces a structured \texttt{JSON} judgment based on a predefined evaluation rubric. Its role is therefore evaluative rather than corrective: it provides a stricter post-hoc assessment of translation quality without altering the translations it audits. In practice, its confidence scores can guide users in prioritizing artifacts for manual inspection, starting with the least confident ones.

The verifier follows several decision criteria:
\begin{enumerate}[nosep, leftmargin=*]
    \item The translated program must genuinely execute the intended benchmark workload rather than bypassing computation through early exits, placeholder implementations, or no-op kernels;
    \item Explicit failure indicators or non-zero error counts in the execution output are treated as definitive failures, regardless of superficial success signals;
    \item Numerical discrepancies are evaluated based on semantic correctness rather than strict textual matching, allowing legitimate floating-point variations;
    \item Surface-level success markers (e.g., ``PASS'') are considered insufficient unless supported by meaningful execution traces and consistent computational results;
    \item Empty, truncated, or otherwise unverifiable outputs are conservatively classified as failures to avoid rewarding silent crashes or incomplete executions.
\end{enumerate}

In addition to a binary pass/fail judgment, the verifier reports a confidence score in the interval $[0,1]$. The confidence score is calibrated using a five-tier rubric that maps the strength of available evidence to numerical ranges:

\begin{enumerate}[nosep, leftmargin=*]
    \item \textbf{0.95--1.0}: The execution output contains explicit verification evidence (e.g., PASS markers, zero failures, or low RMSE), and the translated program is complete;
    \item \textbf{0.8--0.9}: The output appears correct, but the verification evidence is incomplete or partially documented;
    \item \textbf{0.6--0.7}: Execution succeeds, but explicit verification signals are absent and the judgment relies primarily on program analysis;
    \item \textbf{0.4--0.5}: The available evidence is weak or contradictory, making the judgment uncertain;
    \item \textbf{0.0--0.3}: The available information is insufficient for a reliable assessment.
\end{enumerate}

By introducing this LLM-based complementary verification, the framework reduces the risk of overestimating translation quality due to limitations of compiler and runtime checks. The confidence-aware evaluation further encourages conservative judgments in ambiguous cases and provides a more reliable assessment of generated parallel programs.

\section{Evaluation} \label{section:evaluation}

\subsection{Experimental Setup}
\label{subsec:setup}

\paragraph{Benchmarks}
We evaluate our approach on HeCBench~\cite{jin2023hecbench}, a widely-used benchmark collection of HPC and AI kernels written in CUDA, OpenMP, SYCL and HIP. It provides a diverse set of real-world GPU workloads that allow us to assess cross-model code translation in terms of correctness and token usage. In our experiments, we select a representative subset of 30 benchmarks from HeCBench for domain coverage rather than language completeness, spanning a wide range of application domains (Table~\ref{tab:benchmarks}); requiring implementations in all three programming models would skew the subset toward older and simpler kernels. Six of the selected kernels lack OpenMP implementations, leaving 24 benchmarks per OpenMP-source direction.

\paragraph{Models, directions, and baseline}
We conduct experiments with three recent open-source LLMs (DeepSeek-V4-Flash~\cite{deepseek2026v4}, GLM-5.2~\cite{glm52_zai_2026}, and Kimi-K2.7-Code~\cite{kimi2026k27code}) on all six directed language pairs among CUDA, OpenMP, and SYCL. We compare our framework with LASSI~\cite{dearing2024lassi}: to keep the comparison fair, we replace the LLMs used in the original LASSI study (GPT-4, Codestral, WizardCoder, and DeepSeek-Coder-V2) with the same three LLMs, and run both systems on the same benchmark set under identical compilation and execution environments.

\paragraph{Environment and Configuration}
We use \texttt{g++ 8.5.0} for OpenMP target programs, \texttt{nvcc 12.8} for CUDA target programs, and \texttt{icpx/dpcpp 2021.4.0} for SYCL target programs. Compilation is capped at 300 seconds and execution at 120 seconds. The repair loop permits up to 9 iterations.

\begin{table}[t]
    \centering
    \caption{The 30 HeCBench benchmarks used in our evaluation, grouped by domain/pattern.}
    \vspace{-2pt}
    \label{tab:benchmarks}
    \setlength{\tabcolsep}{3pt}
    \renewcommand{\arraystretch}{0.95}
    \small
    \begin{adjustbox}{max width=\columnwidth}
    \begin{tabular}{@{}ll@{}}
        \toprule
        \textbf{Domain} & \textbf{Benchmarks} \\
        \midrule
        Image / computer vision & background-subtract, car, distort, fhd \\
        Deep learning & dense-embedding, p4, pointwise, rowwiseMoments \\
        Math / statistics / trigonometric & atan2, kurtosis, langevin, michalewicz \\
        Sorting / parallel primitives & merge, quicksort, scan, bscan \\
        Stencil / numerical methods & iso2dfd, jacobi \\
        Hashing / cryptography & jenkins-hash, md5hash \\
        Physics / simulation & aidw, md, sheath, tsp \\
        Signal / data processing & fft, interval, reverse2D \\
        RNG / runtime / other & kiss, streamPriority, contract \\
        \bottomrule
    \end{tabular}
    \end{adjustbox}
    \vspace{-8pt}
\end{table}

\begin{table*}[!t]
    \centering
    \caption{Correctness rate and token consumption of our framework and LASSI across three LLMs and six translation directions. For each pair of correctness rates, the higher one is printed in bold green and the lower one in red; token figures are uncolored. The Avg.\ column averages the six directions.}
    \vspace{-2pt}
    \label{tab:correctness-tokens}
    \footnotesize
    \setlength{\tabcolsep}{2pt}
    \renewcommand{\arraystretch}{0.92}
    \begin{tabular*}{\textwidth}{@{\extracolsep{\fill}}llccccccc@{}}
        \toprule
        \multicolumn{9}{c}{\scriptsize correctness (\%) on top / kilo-tokens (translation/repair) below} \\
        \textbf{System} & \textbf{Model}
        & \textbf{CUDA$\!\rightarrow\!$OpenMP} & \textbf{OpenMP$\!\rightarrow\!$CUDA}
        & \textbf{CUDA$\!\rightarrow\!$SYCL} & \textbf{SYCL$\!\rightarrow\!$CUDA}
        & \textbf{OpenMP$\!\rightarrow\!$SYCL} & \textbf{SYCL$\!\rightarrow\!$OpenMP}
        & \textbf{Avg.} \\
        \midrule
        LASSI & DeepSeek V4 Flash
        & \makecell{\lassiC{86.7\%}\\\scriptsize 1327k (59/41)}
        & \makecell{\lassiC{87.5\%}\\\scriptsize 1058k (44/56)}
        & \makecell{\lassiC{46.7\%}\\\scriptsize 1690k (11/89)}
        & \makecell{\lassiC{83.3\%}\\\scriptsize 1174k (48/52)}
        & \makecell{\lassiC{33.3\%}\\\scriptsize 1514k (11/89)}
        & \makecell{\lassiC{53.3\%}\\\scriptsize 1945k (42/58)}
        & \makecell{\lassiC{65.5\%}\\\scriptsize 1451k} \\
        Ours & DeepSeek V4 Flash
        & \makecell{\oursC{93.3\%}\\\scriptsize 718k (72/28)}
        & \makecell{\oursC{95.8\%}\\\scriptsize 835k (53/47)}
        & \makecell{\oursC{70.0\%}\\\scriptsize 1600k (33/67)}
        & \makecell{\oursC{93.3\%}\\\scriptsize 1191k (45/55)}
        & \makecell{\oursC{79.2\%}\\\scriptsize 1325k (35/65)}
        & \makecell{\oursC{100.0\%}\\\scriptsize 753k (73/27)}
        & \makecell{\oursC{88.7\%}\\\scriptsize 1071k} \\
        \midrule
        LASSI & GLM 5.2
        & \makecell{\lassiC{46.7\%}\\\scriptsize 1364k (50/50)}
        & \makecell{\lassiC{60.0\%}\\\scriptsize 741k (55/45)}
        & \makecell{\lassiC{3.3\%}\\\scriptsize 1429k (10/90)}
        & \makecell{\lassiC{53.3\%}\\\scriptsize 2260k (22/78)}
        & \makecell{\lassiC{3.3\%}\\\scriptsize 1258k (10/90)}
        & \makecell{\lassiC{26.7\%}\\\scriptsize 1618k (44/56)}
        & \makecell{\lassiC{34.5\%}\\\scriptsize 1445k} \\
        Ours & GLM 5.2
        & \makecell{\oursC{93.3\%}\\\scriptsize 648k (55/45)}
        & \makecell{\oursC{79.2\%}\\\scriptsize 609k (52/48)}
        & \makecell{\oursC{60.0\%}\\\scriptsize 967k (40/60)}
        & \makecell{\oursC{90.0\%}\\\scriptsize 828k (48/52)}
        & \makecell{\oursC{58.3\%}\\\scriptsize 1042k (32/68)}
        & \makecell{\oursC{86.7\%}\\\scriptsize 2014k (20/80)}
        & \makecell{\oursC{78.6\%}\\\scriptsize 1018k} \\
        \midrule
        LASSI & Kimi K2.7 Code
        & \makecell{\lassiC{56.7\%}\\\scriptsize 3204k (32/68)}
        & \makecell{\lassiC{73.3\%}\\\scriptsize 1088k (58/42)}
        & \makecell{\lassiC{50.0\%}\\\scriptsize 2058k (14/86)}
        & \makecell{\lassiC{93.3\%}\\\scriptsize 1189k (60/40)}
        & \makecell{\lassiC{26.7\%}\\\scriptsize 2731k (10/90)}
        & \makecell{\lassiC{6.7\%}\\\scriptsize 3775k (24/76)}
        & \makecell{\lassiC{54.8\%}\\\scriptsize 2341k} \\
        Ours & Kimi K2.7 Code
        & \makecell{\oursC{100.0\%}\\\scriptsize 860k (88/12)}
        & \makecell{\oursC{100.0\%}\\\scriptsize 810k (77/23)}
        & \makecell{\oursC{93.3\%}\\\scriptsize 1507k (54/46)}
        & \makecell{\oursC{96.7\%}\\\scriptsize 999k (78/22)}
        & \makecell{\oursC{75.0\%}\\\scriptsize 1628k (42/58)}
        & \makecell{\oursC{100.0\%}\\\scriptsize 887k (87/13)}
        & \makecell{\oursC{94.6\%}\\\scriptsize 1115k} \\
        \bottomrule
    \end{tabular*}
    \vspace{-8pt}
\end{table*}

\begin{figure*}[!t]
    \centering
    \subfloat[DeepSeek, OpenMP$\rightarrow$SYCL]{%
        \begin{tikzpicture}
            \begin{axis}[repairbarstop, ymin=400, ymax=1400, ytick={400,800,1200},
                ylabel={Tokens (kilo)}, ylabel style={font=\scriptsize}]
                \addplot coordinates {(0,459.2) (1,661.3) (2,819.7) (3,901.3) (4,986.0) (5,1060.5) (6,1144.7) (7,1204.6) (8,1277.4) (9,1324.9)};
            \end{axis}
            \begin{axis}[repaircurve, ytick=\empty]
                \addplot[repairline]
                    coordinates {(0,8.3) (1,16.7) (2,41.7) (3,45.8) (4,54.2) (5,62.5) (6,75.0) (7,75.0) (8,79.2) (9,79.2)};
            \end{axis}
        \end{tikzpicture}}%
    \subfloat[GLM, OpenMP$\rightarrow$SYCL]{%
        \begin{tikzpicture}
            \begin{axis}[repairbarstop, ymin=280, ymax=1100, ytick={400,700,1000}]
                \addplot coordinates {(0,334.8) (1,493.1) (2,579.6) (3,644.3) (4,705.7) (5,762.5) (6,836.9) (7,906.7) (8,970.7) (9,1041.6)};
            \end{axis}
            \begin{axis}[repaircurve, ytick=\empty]
                \addplot[repairline]
                    coordinates {(0,8.3) (1,20.8) (2,29.2) (3,45.8) (4,54.2) (5,58.3) (6,58.3) (7,58.3) (8,58.3) (9,58.3)};
            \end{axis}
        \end{tikzpicture}}%
    \subfloat[Kimi, OpenMP$\rightarrow$SYCL]{%
        \begin{tikzpicture}
            \begin{axis}[repairbarstop, ymin=600, ymax=1700, ytick={700,1100,1500}]
                \addplot coordinates {(0,676.1) (1,883.6) (2,1012.2) (3,1105.7) (4,1203.7) (5,1307.8) (6,1383.0) (7,1461.0) (8,1564.2) (9,1627.5)};
            \end{axis}
            \begin{axis}[repaircurve, ytick={0,50,100},
                ylabel={Success rate (\%)}, ylabel style={font=\scriptsize}]
                \addplot[repairline]
                    coordinates {(0,12.5) (1,29.2) (2,45.8) (3,54.2) (4,62.5) (5,66.7) (6,75.0) (7,75.0) (8,75.0) (9,75.0)};
            \end{axis}
        \end{tikzpicture}}%
    \\[0pt]
    \subfloat[DeepSeek, CUDA$\rightarrow$OpenMP]{%
        \begin{tikzpicture}
            \begin{axis}[repairbarsx, ymin=450, ymax=750, ytick={500,600,700},
                ylabel={Tokens (kilo)}, ylabel style={font=\scriptsize}]
                \addplot coordinates {(0,518.2) (1,598.4) (2,619.4) (3,634.1) (4,644.5) (5,658.3) (6,670.7) (7,683.6) (8,708.7) (9,718.2)};
            \end{axis}
            \begin{axis}[repaircurve, ytick=\empty]
                \addplot[repairline]
                    coordinates {(0,56.7) (1,70.0) (2,83.3) (3,90.0) (4,93.3) (5,93.3) (6,93.3) (7,93.3) (8,93.3) (9,93.3)};
            \end{axis}
        \end{tikzpicture}}%
    \subfloat[DeepSeek, OpenMP$\rightarrow$CUDA]{%
        \begin{tikzpicture}
            \begin{axis}[repairbarsx, ymin=380, ymax=880, ytick={400,600,800}]
                \addplot coordinates {(0,445.6) (1,628.2) (2,694.5) (3,727.9) (4,738.4) (5,748.2) (6,766.9) (7,791.6) (8,816.9) (9,835.3)};
            \end{axis}
            \begin{axis}[repaircurve, ytick=\empty]
                \addplot[repairline]
                    coordinates {(0,33.3) (1,66.7) (2,83.3) (3,95.8) (4,95.8) (5,95.8) (6,95.8) (7,95.8) (8,95.8) (9,95.8)};
            \end{axis}
        \end{tikzpicture}}%
    \subfloat[DeepSeek, CUDA$\rightarrow$SYCL]{%
        \begin{tikzpicture}
            \begin{axis}[repairbarsx, ymin=480, ymax=1650, ytick={600,1000,1400}]
                \addplot coordinates {(0,534.3) (1,758.7) (2,903.7) (3,1019.7) (4,1125.4) (5,1210.9) (6,1294.5) (7,1411.4) (8,1509.8) (9,1599.9)};
            \end{axis}
            \begin{axis}[repaircurve, ytick={0,50,100},
                ylabel={Success rate (\%)}, ylabel style={font=\scriptsize}]
                \addplot[repairline]
                    coordinates {(0,10.0) (1,23.3) (2,43.3) (3,60.0) (4,70.0) (5,70.0) (6,70.0) (7,70.0) (8,70.0) (9,70.0)};
            \end{axis}
        \end{tikzpicture}}%
    \caption{Repair loop effectiveness in six representative model and direction settings. In each subfigure, bars (left axis) report the cumulative token consumption in kilo-tokens and the curve (right axis) reports the cumulative success rate. Note that the left axes are truncated and do not start from zero. Round 0 is the initial translation before any repair; rounds 1 to 9 are repair invocations. Top row: OpenMP$\rightarrow$SYCL, the most difficult direction in Table~\ref{tab:correctness-tokens}, under all three models; bottom row: DeepSeek V4 Flash under CUDA$\rightarrow$OpenMP, OpenMP$\rightarrow$CUDA, and CUDA$\rightarrow$SYCL.}
    \label{fig:repair-rounds}
    \vspace{-5pt}
\end{figure*}
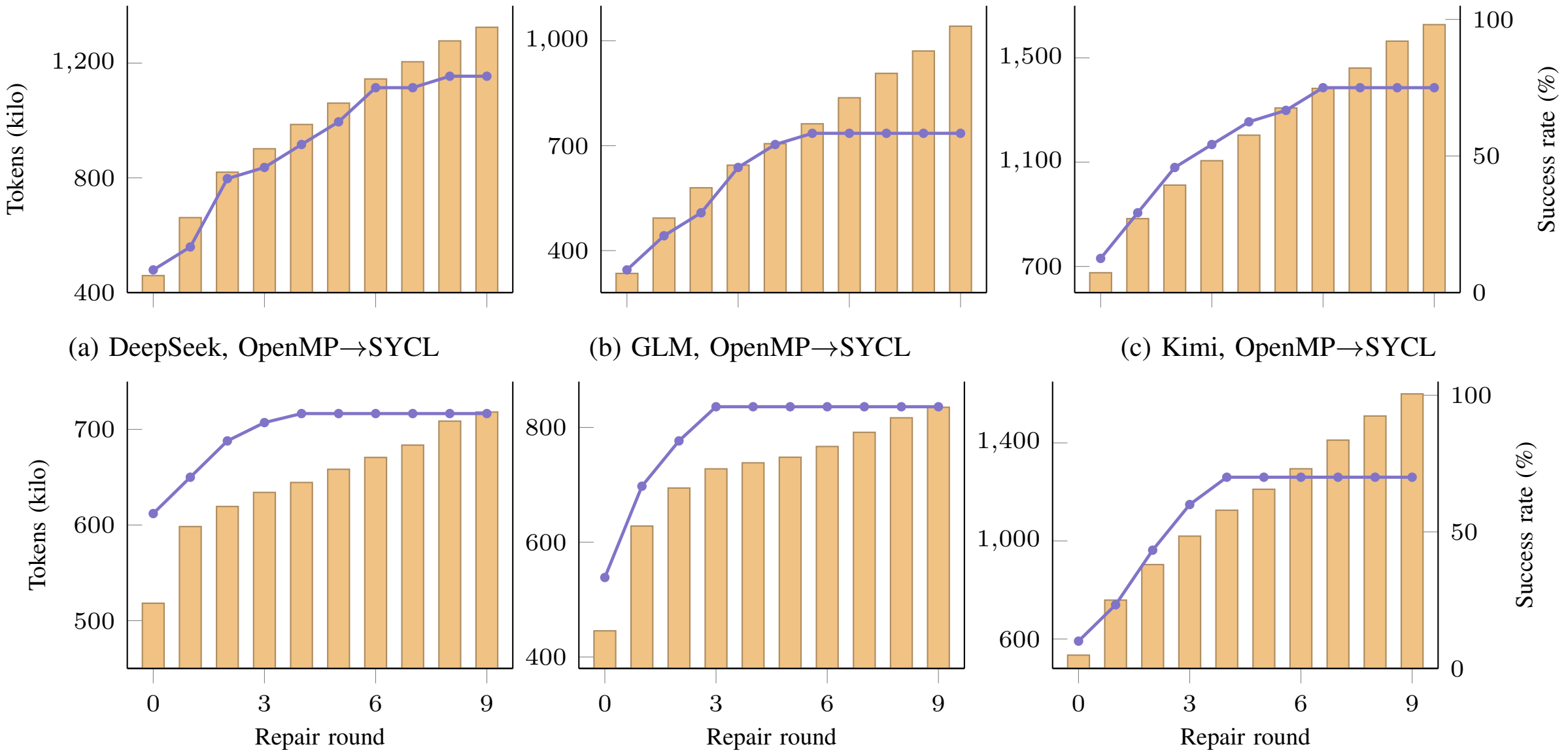

\subsection{Correctness and Token Consumption Evaluation among OpenMP, CUDA, and SYCL}
Table~\ref{tab:correctness-tokens} reports, for every system, model, and direction, the verified correctness rate together with the total token consumption and its translation and repair split. We start with DeepSeek V4 Flash on the CUDA$\leftrightarrow$OpenMP pair, the most common translation scenario in HPC practice and also the pair LASSI was originally designed for. Both systems perform well here: our framework translates 93.3\% (CUDA$\rightarrow$OpenMP) and 95.8\% (OpenMP$\rightarrow$CUDA) of the benchmarks correctly, against LASSI's 86.7\% and 87.5\%. The two systems differ mainly in how much they spend and where the tokens go. LASSI consumes 1,327k and 1,058k tokens on the two directions, of which 41\% and 56\% go into repairing retrieval-based drafts. Our framework consumes 718k and 835k tokens, 45.9\% and 21.1\% less, with only 28\% and 47\% spent on repair. 

Beyond the CUDA$\leftrightarrow$OpenMP pair LASSI was designed for, we add SYCL as a representative low-resource parallel programming model and extend the evaluation to all six directions with the same model. Our framework attains the higher correctness rate on every direction and averages 88.7\% against LASSI's 65.5\%. The gap is smallest on SYCL$\rightarrow$CUDA (93.3\% vs.\ 83.3\%), where both systems can rely on the model's familiarity with CUDA-style code, and largest on OpenMP$\rightarrow$SYCL (79.2\% vs.\ 33.3\%), where LASSI's retrieval corpus offers little reusable material. Runs with SYCL as the target are intrinsically harder for both systems, averaging 74.6\% for our framework and 40.0\% for LASSI against 96.7\% and 68.3\% with SYCL as the source, while the CUDA$\leftrightarrow$OpenMP pair is nearly symmetric (93.3\% vs.\ 95.8\%). Both systems read low-resource SYCL far better than they write it. The average token consumption is 26.2\% lower at the same time (1,071k vs.\ 1,451k per direction). The only direction where our framework spends more is SYCL$\rightarrow$CUDA (1,191k vs.\ 1,174k, +1.5\%), and there the extra repair rounds come with a 10.0\% correctness gain. Beyond these direction-level trends, the failing benchmarks show no obvious pattern, scattering across domains and kernel styles.

\subsection{Generalization Across Models}
The remaining rows of Table~\ref{tab:correctness-tokens} repeat the experiment with GLM-5.2 and Kimi-K2.7-Code, and the key observation is that the performance of our framework is largely agnostic to the choice of LLM. Its per-model averages span only 16.0\% (78.6\% to 94.6\%), and every direction stays from 6.7\% to 33.3\% across models (17.0\% on average), because the verify-and-repair loop compensates for weaker first drafts instead of relying on a particular model's strengths. LASSI, by comparison, fluctuates far more: 31.0\% across its per-model averages and 38.5\% per direction, with swings of 27.5\% to 40.0\% even on its three strongest directions. Our framework also attains the higher correctness rate in all 18 combinations. Token consumption is equally predictable, staying within 1,018k to 1,115k tokens per direction (a spread below 10\%; LASSI: 1,445k to 2,341k), 29.5\% to 52.4\% below LASSI per model; only two of the 18 cells spend more than LASSI, and both buy large correctness gains (up to +60.0\% for +24.5\% tokens).

\subsection{Repair Loop Effectiveness Evaluation}
To isolate the contribution of the repair loop, Figure~\ref{fig:repair-rounds} follows six representative model and direction combinations through the loop, plotting the cumulative token consumption (bars, left axis) and the cumulative success rate (curve, right axis) after each round, where round 0 is the initial translation before any repair. The top row focuses on the hardest direction in Table~\ref{tab:correctness-tokens}, OpenMP$\rightarrow$SYCL, whose three-model averages are the lowest for both systems (70.8\% for our framework and 21.1\% for LASSI), and shows it under all three models. All three start at or below 12.5\%, and the success rate climbs steadily as the loop accumulates tokens, reaching 79.2\%, 58.3\%, and 75.0\% within nine repairs: gains of 50.0\% to 70.9\% that come entirely from verification-guided repair rather than from the first draft. The bottom row varies the direction for DeepSeek V4 Flash. On the easier CUDA$\leftrightarrow$OpenMP pair the strong first drafts (56.7\% and 33.3\% solved before any repair) explain the pair's low token consumption in Table~\ref{tab:correctness-tokens}, yet the loop still adds 36.6\% and 62.5\% to reach 93.3\% and 95.8\%; on CUDA$\rightarrow$SYCL it adds 60.0\% to reach 70.0\%.
In all six panels the success rate rises monotonically with the accumulated tokens, showing that the repair loop, not the initial generation, accounts for most of the final correctness.

\begin{table}[!t]
    \centering
    \caption{Pass rates (\%) under the simple in-loop verification (In-loop) and the LLM-based complementary verification (Compl.\ verif.). $\Delta$ is the change in percentage points; FP counts artifacts passed by the former but rejected by the latter. No false negatives were found. Nine LASSI runs with Kimi-K2.7-Code produced no output and are excluded.}
    \vspace{-3pt}
    \label{tab:audit}
    \small
    \setlength{\tabcolsep}{2pt}
    \begin{tabular}{@{}llccccr@{}}
        \toprule
        \textbf{System} & \textbf{Model} & \textbf{$N$} & \textbf{In-loop} & \makecell{\textbf{Compl.}\\\textbf{verif.}} & $\Delta$ & \textbf{FP} \\
        \midrule
        Ours  & DeepSeek V4 Flash & 168 & 88.7 & 71.4 & $-$17.3 & 29 \\
        Ours  & GLM 5.2           & 168 & 78.6 & 68.5 & $-$10.1 & 17 \\
        Ours  & Kimi K2.7 Code    & 168 & 94.6 & 86.3 & $-$8.3  & 14 \\
        \midrule
        LASSI & DeepSeek V4 Flash & 168 & 65.5 & 46.4 & $-$19.1 & 32 \\
        LASSI & GLM 5.2           & 168 & 34.5 & 28.0 & $-$6.5  & 11 \\
        LASSI & Kimi K2.7 Code    & 159 & 57.9 & 49.1 & $-$8.8  & 14 \\
        \midrule
        \multicolumn{2}{@{}l}{Overall} & 999 & 70.1 & 58.4 & $-$11.7 & 117 \\
        \bottomrule
    \end{tabular}
    \vspace{-5pt}
\end{table}

\subsection{LLM-based Complementary Verification}
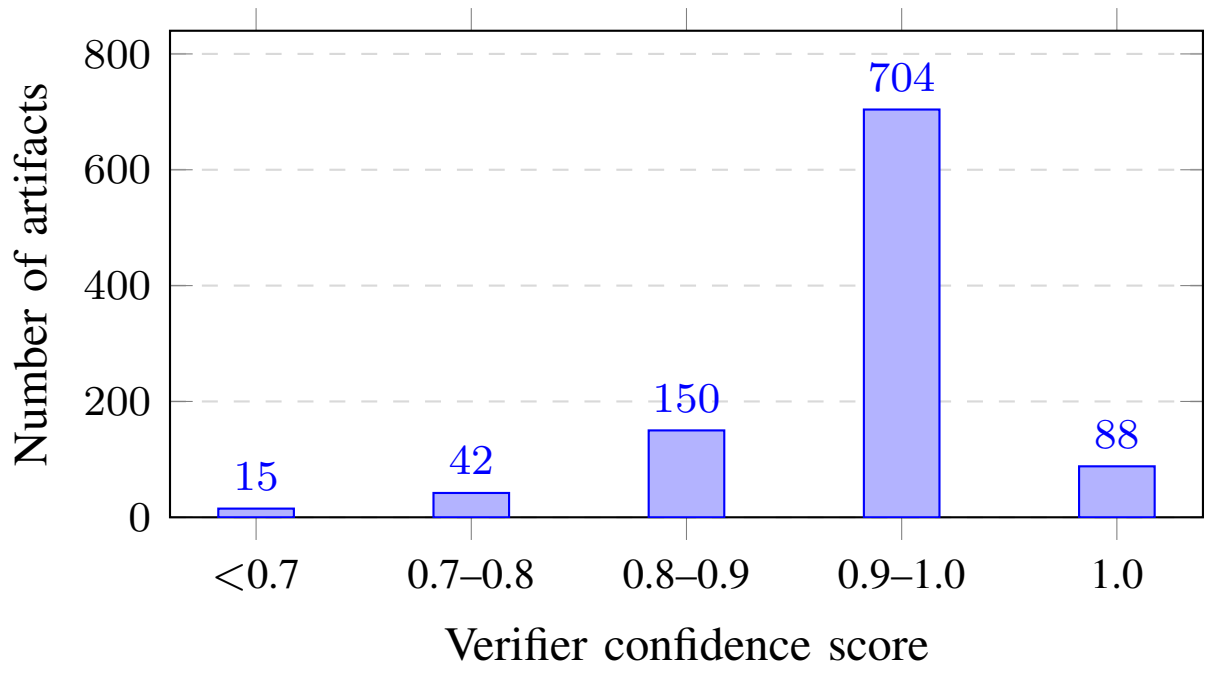
\begin{figure}[t]
    \centering
    \begin{tikzpicture}
        \begin{axis}[
            width=0.95\linewidth, height=4.8cm, ybar, bar width=0.5cm,
            xlabel={Verifier confidence score}, ylabel={Number of artifacts},
            symbolic x coords={$<$0.7, 0.7--0.8, 0.8--0.9, 0.9--1.0, 1.0},
            xtick=data, ymin=0, ymax=840,
            nodes near coords, nodes near coords style={font=\footnotesize},
            every axis plot/.append style={fill=blue!55, draw=blue!70!black},
            tick label style={font=\footnotesize},
            xlabel style={font=\small}, ylabel style={font=\small},
            ymajorgrids, grid style={dashed, gray!30},
        ]
            \addplot coordinates {($<$0.7, 15) (0.7--0.8, 42) (0.8--0.9, 150) (0.9--1.0, 704) (1.0, 88)};
        \end{axis}
    \end{tikzpicture}
    \vspace{-8pt}
    \caption{Distribution of the verifier's confidence scores over the 999 artifacts checked by the complementary verification. The median confidence is 0.95, and 79.3\% of the judgments carry a confidence score of at least 0.9.}
    \label{fig:audit-confidence}
    \vspace{-8pt}
\end{figure}

All success rates reported so far are judged by the simple in-loop verification stage, which compiles each artifact, runs it, and checks its output. To assess how much these judgments can be trusted, we additionally applied the rigorous LLM-based complementary verification (Section~\ref{sec:verification}), backed by DeepSeek V4 Flash, to the cached artifacts of both systems. The verifier issues a pass or fail judgment together with a confidence score; Figure~\ref{fig:audit-confidence} shows the confidence distribution over the 999 artifacts that produced output (1,008 runs in total, minus 9 LASSI runs with Kimi K2.7 Code where, due to the complexity of the tasks, the model repeatedly failed to complete its output within the API time limit).

\begin{table*}[!t]
    \centering
    \caption{Taxonomy of the 117 false positives identified by the LLM-based complementary verification, with one representative case per escape mode. All examples are drawn from DeepSeek V4 Flash runs, which contribute the largest share of false positives.}
    \vspace{-2pt}
    \label{tab:audit-cases}
    \footnotesize
    \setlength{\tabcolsep}{4pt}
    \renewcommand{\arraystretch}{0.95}
    \begin{tabular}{@{}ll>{\raggedright\arraybackslash}p{3.9cm}>{\raggedright\arraybackslash}p{6.4cm}@{}}
        \toprule
        \textbf{Escape mode} & \textbf{Count (\%)} & \textbf{Representative case} & \textbf{Key evidence} \\
        \midrule
        Placeholder / stub implementation & 31 (26.5) & LASSI, CUDA$\rightarrow$OpenMP, contract   & All-zero checksums; the kernel only zero-fills the output \\
        Numerical error                   & 28 (23.9) & LASSI, OpenMP$\rightarrow$CUDA, distort    & Per-channel maximum error of 255 \\
        Verification skipped              & 25 (21.4) & LASSI, CUDA$\rightarrow$OpenMP, pointwise  & \texttt{PASS} printed with reference validation removed \\
        Parameter mismatch                & 12 (10.3) & Ours, SYCL$\rightarrow$CUDA, distort       & Hard-coded $K=0.3$ while the test argument is $0.001$ \\
        Index / buffer / memory error     & 9 (7.7)   & LASSI, OpenMP$\rightarrow$CUDA, car        & RMSE of 61{,}218; channel dimension ignored \\
        RNG / seed issue                  & 4 (3.4)   & Ours, SYCL$\rightarrow$OpenMP, kiss        & One RNG seed per thread instead of one per element \\
        Kernel not executed               & 2 (1.7)   & LASSI, CUDA$\rightarrow$SYCL, reverse2D    & Kernel execution time reported as \texttt{-nan} \\
        Other                             & 6 (5.1)   & --- & --- \\
        \bottomrule
    \end{tabular}
    \vspace{-8pt}
\end{table*}

Table~\ref{tab:audit} compares the pass rates under the two verification stages. The rigorous complementary verification lowers the pass rates of both systems, by 8.3\% to 17.3\% for our framework and by 6.5\% to 19.1\% for LASSI, and it finds no false negatives: no artifact rejected in the loop is accepted by the verifier. Under the stricter check our framework still passes 71.4\%, 68.5\%, and 86.3\% of its artifacts with the three models, keeping margins of 25.0\%, 40.5\%, and 37.2\% over LASSI. In total, 117 artifacts (11.7\%) are false positives of the in-loop stage, and Table~\ref{tab:audit-cases} groups them into eight escape modes led by placeholder implementations (31 cases), numerical errors (28), and skipped verification (25). The verifier is confident in most of its judgments: the median confidence is 0.95, 79.3\% of the judgments score at least 0.9, and only 57 artifacts (5.7\%) fall below 0.8, so manual spot-checking of the uncertain remainder stays feasible. Combining the stricter pass rates with token consumption, our framework spends 46k to 54k tokens per verified successful benchmark across the three models, against 112k to 184k for LASSI, a 2.1 to 3.9 times advantage in cost per verified success.

\section{Discussion}
\label{section:discussion}

The experiments and the design process itself yield the following findings about LLM-based translation of parallel programs under verification-guided repair.

\begin{enumerate}[nosep, leftmargin=*]

    \item \textbf{Skills make both translation and repair more token-efficient.}
    Analysis and memory-planning skills target translation at an explicit parallelization plan rather than a loosely matching retrieved example, and verification-classified feedback targets repair at the diagnosed defect rather than blind regeneration, keeping the repair share of the budget smaller across all models and directions.

    \item \textbf{Translation difficulty is determined by the target model, not the source.}
    Directions with the same target cluster together in correctness, with low-resource SYCL consistently the hardest target: current LLMs read low-resource parallel code far better than they write it, which is also why retrieval offers little help on such directions.

    \item \textbf{The repair loop accounts for most of the final correctness, but returns diminish beyond roughly five rounds.}
    Success climbs monotonically from a low round-0 base on hard directions and still adds large gains on easy pairs, with most gains arriving within five rounds; later rounds mostly accumulate tokens. First-draft strength determines what a direction costs, not how many benchmarks eventually pass, and failures that survive many rounds are systematic ones that generous iteration caps cannot fix.

    \item \textbf{Verification and repair make the framework largely model-agnostic.}
    Correctness and token consumption stay in narrow bands across models while the retrieval-based baseline swings widely on both; the loop compensates for weaker first drafts, so framework-level mechanisms matter more than the choice of LLM.

    \item \textbf{Compiler and runtime checks are necessary but not sufficient.}
    In-loop verification overestimates the success of both systems, with systematic escapes led by placeholder implementations, weakened validation, and silent numerical errors. The complementary verification adds no false negatives and preserves the margins between systems, and its calibrated confidence concentrates at the high end, leaving only a small uncertain tail for manual spot-checking.

    \item \textbf{Separating generation, failure classification, and repair keeps self-correction bounded.}
    Verdict routing gives each repair skill a single failure mode and the context it needs, and the unified budget prevents runaway retries without penalizing clean runs; the same encapsulation makes the framework direction-agnostic.

\end{enumerate}

\section{Conclusion and Future Work}
\label{section:conclusion}

This paper presented a multi-agent system for parallel scientific code translation, motivated by two gaps in existing pipelines: they verify correctness against criteria that do not match scientific computing practice, and they spend token budgets on blind, undifferentiated retries. Our framework separates translation, compilation diagnosis, and execution validation into three agents coordinated by a verdict-driven orchestrator, routes each diagnosed failure to a targeted repair skill, and treats automated output checking as a first-class validation step. On 30 HeCBench kernels translated bidirectionally among OpenMP, CUDA, and SYCL with three recent open-source LLMs, our framework attains the higher verified correctness rate in all 18 model and direction combinations over LASSI, with per-model averages of 78.6\% to 94.6\% against 34.5\% to 65.5\%, while consuming 29.5\% to 52.4\% fewer tokens. Under the stricter LLM-based complementary verification, it preserves margins of 25.0\% to 40.5\% at a 2.1 to 3.9 times advantage in cost per verified success. 
The experiments further show that the repair loop accounts for most of the final correctness; verification-guided repair makes the framework largely model-agnostic; and compiler and runtime checks alone systematically overestimate translation success.
More capable translation behavior, in other words, comes not from more expensive prompting but from better verification scaffolding around the model.

Several challenges remain open. First, our evaluation is limited to kernel-level benchmarks; extending the framework to multi-file codebases and full applications will require cross-file analysis, build-system awareness, and translation state that persists across compilation units. Second, the LLM-based complementary verification can itself be strengthened, for example through refined evaluation rubrics, multi-verifier voting, and tighter coupling between confidence scores and targeted re-execution, so that the uncertain tail requiring manual inspection shrinks further. Third, correctness is only the first step toward practical porting: we plan to incorporate performance feedback into the repair loop so that translated programs are not only numerically correct but also performant, and to extend the direction-agnostic skill mechanism to further programming models such as HIP and Kokkos, where the low-resource difficulty observed for SYCL targets is likely to be even more pronounced.


\section*{Acknowledgment}
This research used the DeepBlizzard GPU HPC cluster at Michigan Technological University, which was funded by NSF Major Research Instrumentation (MRI) Grant Award Number 2215734, MRI: Acquisition of a GPU-accelerated cluster for research, training, and outreach.




\bibliographystyle{IEEEtran}

\bibliography{bibliography.bib}
%
\clearpage


\twocolumn[%
{\begin{center}
\Huge
Appendix: Artifact Description/Artifact Evaluation
\end{center}}
]


\appendixAD

\section{Overview of Contributions and Artifacts}

\subsection{Paper's Main Contributions}

\begin{description}
\item[$C_1$] A multi-agent framework for automated HPC code translation across
CUDA, OpenMP, and SYCL, comprising a TranslationAgent, a CompilationAgent,
and an ExecutionAgent coordinated by a verdict-driven Orchestrator.
\item[$C_2$] A set of repair skills (Patch, Semantic, Sync, Memory) that detect
and fix compilation, runtime, and correctness errors during translation,
enabling iterative self-correction without human intervention.
\item[$C_3$] An evaluation of the framework on 30 HeCBench benchmarks across
six translation directions and three LLM backends (DeepSeek V4 Flash, GLM 5.2,
Kimi K2.7 Code), demonstrating higher verified correctness and lower token
consumption than the LASSI baseline.
\end{description}

\subsection{Computational Artifacts}

\begin{description}
\item[$A_1$] \url{https://github.com/dinglhmds/More-with-Less-Efficient-Parallel-Scientific-Code-Translation-via-an-Agent-Skill-Workflow}
\end{description}

\begin{center}
\begin{tabular}{rll}
\toprule
Artifact ID  &  Contributions &  Related \\
             &  Supported     &  Paper Elements \\
\midrule
$A_1$   &  $C_1,C_2,C_3$ & Tables IV--VI \\
        &                & Figures 2--3 \\
\bottomrule
\end{tabular}
\end{center}

\section{Artifact Identification}

\newartifact

\artrel

The artifact implements the full multi-agent translation pipeline described in
$C_1$ and $C_2$, including the Orchestrator, all three agents, all repair
skills, and the offline LLM-based complementary verifier. It also contains
the LASSI baseline (configured with the same three LLMs), the batch runner,
and the evaluation scripts used to produce the results in $C_3$.

\artexp

Running the batch evaluation on the 30 HeCBench benchmarks (six translation
directions, three LLMs, both systems) should produce per-benchmark verdicts
(success/failure), token consumption reports, cumulative success-rate curves,
and complementary-verification judgments matching Tables IV--VI and
Figures 2--3 in the paper. The framework is expected to achieve a higher
verified correctness rate than the LASSI baseline while consuming fewer LLM
tokens on average.

\arttime

Artifact Setup: approximately 30 minutes (install Python dependencies and
configure LLM API access). Artifact Execution: approximately 8--12 hours for
all 30 benchmarks on a GPU cluster (dominated by LLM API latency). Artifact
Analysis: approximately 15 minutes (aggregate per-benchmark result.json files
into summary tables and plots).

\artin

\artinpart{Hardware}

A GPU cluster managed by Slurm with at least one NVIDIA GPU node is required
for executing CUDA baselines and OpenMP/SYCL offload targets. CPU-only
OpenMP execution is also supported as a fallback. The evaluation used a
Slurm cluster with the \texttt{mrigpu} partition (one GPU per job).

\artinpart{Software}

\begin{itemize}
    \item Python 3.9 or newer: \url{https://www.python.org}
    \item OpenAI Python client ($\geq$1.0): \url{https://github.com/openai/openai-python}
    \item PyYAML ($\geq$6.0): \url{https://pyyaml.org}
    \item python-dotenv ($\geq$1.0): \url{https://github.com/theskumar/python-dotenv}
    \item g++ 8.5.0: \url{https://gcc.gnu.org}
    \item CUDA Toolkit 12.8 (nvcc): \url{https://developer.nvidia.com/cuda-toolkit}
    \item Intel oneAPI DPC++/C++ Compiler 2021.4.0 (icpx/dpcpp): \url{https://www.intel.com/content/www/us/en/developer/tools/oneapi/dpc-compiler.html}
    \item Clang/LLVM 18.1.8 (optional, for OpenMP GPU offload): \url{https://clang.llvm.org}
    \item Access to an OpenAI-compatible LLM API (DeepSeek, GLM, Kimi, or equivalent)
\end{itemize}

\artinpart{Datasets / Inputs}

The evaluation uses 30 benchmarks selected from HeCBench
(\url{https://github.com/zjin-lcf/HECBench}), covering image processing,
deep learning, mathematical and statistical kernels, sorting and parallel
primitives, stencil and numerical methods, hashing, physics simulation, and
signal processing. The selected benchmark names are listed in
\texttt{DEFAULT\_BENCHMARKS} in \texttt{batch\_runner.py}; the suite itself is
obtained by cloning the HeCBench repository and passing its \texttt{src}
directory to the batch runner via \texttt{-\/-benchmark-dir}. Six of the
selected kernels lack OpenMP implementations, leaving 24 benchmarks per
OpenMP-source direction. No other external datasets are required.

\artinpart{Installation and Deployment}

Clone the repository, create a Python virtual environment, and install
dependencies:

\begin{verbatim}
python3 -m venv .venv
source .venv/bin/activate
pip install -r requirements.txt
\end{verbatim}

Configure the LLM API endpoint by copying \texttt{.env.example} to
\texttt{.env} and setting \texttt{BASE\_URL}, \texttt{API\_KEY}, and
\texttt{MODEL\_NAME}.

Set environment variables for compilation and execution:

\begin{verbatim}
export CT_CUDA_PATH=/path/to/cuda/12.8
# optional, for OpenMP GPU offload:
export CT_CLANG_HOME=/path/to/llvm/18.1.8
export LD_LIBRARY_PATH=\
    $CT_CLANG_HOME/lib:$LD_LIBRARY_PATH
export CT_SLURM_PARTITION=mrigpu
export CT_SLURM_GRES=gpu:1
\end{verbatim}

\artcomp

The experiment workflow consists of four tasks:
$T_1$ (translation) $\rightarrow$ $T_2$ (compilation/execution)
$\rightarrow$ $T_3$ (result aggregation) $\rightarrow$ $T_4$ (offline
complementary verification). The full evaluation matrix comprises two
systems (our framework and the LASSI baseline), three LLMs, and six
translation directions; individual runs are independent and can be
distributed across Slurm jobs.

$T_1$: The batch runner (\texttt{batch\_runner.py}) discovers benchmarks
under the HeCBench source directory, reads each source program, and runs
the multi-agent pipeline. The Orchestrator coordinates Analyzer,
MemoryPlanning, and Translator skills to produce target code, then invokes
the CompilationAgent and the ExecutionAgent. If compilation or execution
fails, repair skills mutate the code and the pipeline re-enters from the
corresponding stage. Each benchmark is capped at 300 seconds for
compilation, 120 seconds for execution, and a maximum of 9 repair
iterations.

$T_2$: Each source program is compiled with its native compiler
(\texttt{nvcc -O3 -std=c++17} for CUDA, \texttt{g++ -O2 -fopenmp -std=c++17}
for OpenMP, \texttt{icpx -fsycl -std=c++17} for SYCL) and executed on a
Slurm GPU node to obtain reference outputs. Translated programs are compiled
with the corresponding target compiler (for OpenMP GPU offload,
\texttt{clang++ -O3 -fopenmp -fopenmp-targets=nvptx64 -std=c++17}) and
executed with the same input arguments.

$T_3$: Per-benchmark results are written to
\texttt{output/<timestamp>/\allowbreak<benchmark>/result.json}, containing the
verdict, retry counts, elapsed time, and token usage broken down by category
and repair round. A summary script aggregates these files into the tables
and figures reported in the paper.

$T_4$: The offline LLM-based complementary verifier is applied to the cached
artifacts of both systems. Given the complete source program and the
execution outputs of a translated program, it produces a structured
\texttt{JSON} judgment (pass/fail plus a confidence score in $[0,1]$)
following the rubric described in the paper. Aggregating these judgments
yields the complementary-verification pass rates in Table~V, the
false-positive taxonomy in Table~VI, and the confidence distribution in
Figure~3.

\artout

The expected output is a set of per-benchmark \texttt{result.json} files,
per-artifact verification judgments, and a summary report. The key metrics
are: (1) the verified correctness rate (the \texttt{EQUIVALENT} verdict,
routed to the terminal Done state), (2) token consumption aggregated by
direction and model, (3) the cumulative success rate as a function of repair
rounds, and (4) the complementary-verification pass rates and confidence
distribution. These correspond to Tables IV--VI and Figures 2--3 in the
paper.

\newpage
\appendixAE

\arteval{1}

\artin

\textbf{Access.} The artifact is available at
\url{https://github.com/dinglhmds/More-with-Less-Efficient-Parallel-Scientific-Code-Translation-via-an-Agent-Skill-Workflow}.
Reviewers are granted access to an OpenAI-compatible LLM endpoint; the
endpoint URL and an API key with sufficient quota are provided through the
submission system.

\textbf{Hardware.} One NVIDIA GPU node on a Slurm-managed cluster (one GPU
per job), as described in the AD Appendix.

\textbf{Software.} Python 3.9+ with the packages listed in
\texttt{requirements.txt} (OpenAI client, PyYAML, python-dotenv); g++ 8.5.0,
CUDA Toolkit 12.8, and Intel oneAPI DPC++/C++ Compiler 2021.4.0. Exact
installation commands and environment variables are given in the AD Appendix
(Artifact Setup) and in the repository \texttt{README.md}.

\artcomp

To stay within the evaluation time budget, we define a reduced but
representative evaluation: the CUDA$\rightarrow$OpenMP direction with
DeepSeek V4 Flash, all 30 benchmarks, run for both our framework and the
LASSI baseline. This reproduces one column pair of Table~IV and the
bottom-left panel of Figure~2. The model is selected via
\texttt{MODEL\_NAME} in \texttt{.env}.

$E_1$ (translation with self-correction): for our framework, run
\begin{verbatim}
export CT_TRANSLATION_DIRECTION=\
    cuda_to_openmp
python batch_runner.py \
    --benchmark-dir <hecbench>/src \
    --output batch_cuda2omp_ae.json \
    --no-context --no-rag \
    --use-memory-plan \
    --auto-scale-on-timeout \
    --max-retries 5
\end{verbatim}
and for the LASSI baseline, run
\begin{verbatim}
python experiments/lassi/\
    run_lassi30_cuda2omp_context.py
\end{verbatim}
(On a Slurm cluster, the equivalent submission scripts are
\path{scripts/submit_framework_cuda_to_openmp_norag_nocontext.sh}
and \path{scripts/submit_lassi_cuda2omp_context.sh}; adapt the
cluster-specific paths before use.) For each benchmark this executes tasks
$T_1$ and $T_2$ of the AD workflow: the verdict-driven translation loop,
compilation with \texttt{g++ -O2 -fopenmp -std=c++17}, and execution on the
GPU node, capped at 300~s compilation, 120~s execution, and 9 repair
iterations.

$E_2$ (aggregation): run
\begin{verbatim}
python tools/\
    framework_model_direction_summary.py
python tools/framework_round_summary.py
\end{verbatim}
which aggregate all \texttt{result.json} files into a summary table and the
cumulative success-rate data (task $T_3$).

Optionally, $E_3$ (complementary verification): run
\begin{verbatim}
python tools/run_llm_verdict_full.py
\end{verbatim}
to apply the offline LLM-based verifier (task $T_4$) to the cached artifacts,
producing per-artifact pass/fail judgments with confidence scores.

\artout

\textbf{Expected results.} $E_1$ produces 30 \texttt{result.json} files per
system; $E_2$ prints a table with the verified correctness rate and total
token consumption per system, together with the cumulative success-rate curve
over repair rounds.

\textbf{Evaluation methodology.} Compare the summary table against the
CUDA$\rightarrow$OpenMP columns of Table~IV: our framework should achieve a
higher verified correctness rate than LASSI (paper: 93.3\% vs.\ 86.7\%) at
lower token consumption (paper: 718k vs.\ 1,327k). Because LLM outputs are
nondeterministic, absolute rates may shift by a few percentage points between
runs; the reproduction criterion is that the ordering of the two systems and
the approximate split of tokens between translation and repair are preserved.
The cumulative curve should increase monotonically, matching the bottom-left
panel of Figure~2.

\textbf{Correlation with contributions.} Successful completion of the
verdict-driven loop, including repair invocations, substantiates $C_1$ and
$C_2$; the correctness and token comparison between the two systems
substantiates $C_3$. If $E_3$ is executed, the verifier's judgments
additionally provide the stricter complementary pass rates underlying
Table~V.
\end{document}